# Shot Noise Suppression in Avalanche Photodiodes

Feng Ma, Shuling Wang, and Joe C. Campbell[a)]

*Microelectronics Research Center, Department of Electrical and Computer Engineering, the University of Texas at Austin, Austin, TX 78712*

## Abstract

We identify a new shot noise suppression mechanism in a thin (~100 nm) heterostructure avalanche photodiode. In the low-gain regime the shot noise is suppressed due to temporal correlations within amplified current pulses. We demonstrate in a Monte Carlo simulation that the effective excess noise factors can be <1, and reconcile the apparent conflict between theory and experiments. This shot noise suppression mechanism is independent of known mechanisms such as Coulomb interaction, or reflection at heterojunction interfaces.

[a)] Author to whom correspondence should be addressed; electronic mail: jcc@mail.utexas.edu

Shot noise suppression in mesoscopic devices has drawn a lot of attention in recent years, as noise contains ample information of physical processes. González et al. found a universal shot noise suppression factor of 1/3 in non-degenerate diffusive conductors, a result of elastic electron scattering [1]. Oberholzer et al. studied [2] the partitioning of the electron wave diffraction inside a chaotic cavity that results in a Fano factor of 1/4. Strong shot noise suppression has been observed in ballistic quantum point contacts, due to temporally correlated electrons, possibly a consequence of space-charge effect (Coulomb interactions) [3]. Shot noise is also suppressed when low-energy (<0.3 eV) electrons resonant-tunnel through two barriers made of $Al_xGa_{1-x}As$, with a GaAs quantum well between the barriers [4]. Other causes of shot noise suppression include Fermi-Dirac statistics through the Pauli principle [5] and phase coherent transport [6]. All these examples involve shot noise *suppression* because there is no gain mechanism involved. On the other hand in a resonant-tunneling diode, biased in the negative differential resistance regime, a tunneled electron raises the potential energy of a quantum well producing more available states for more tunneling. Such a *positive* correlation results in a shot noise *enhancement* [7].

Multiplication noise in avalanche photodiodes (APDs) has been studied extensively in the literature [8-13], and a common figure of merit is the so called excess noise factor, $F(<M>)=<M^2>/<M>^2$, as a function of the mean gain $<M>$. By this definition $F(<M>)\geq 1$. However, recently Wang et al. [14] have measured excess noise factors < 1 (Fig. 1) at low gains for an Impact-Ionization-Engineered ($I^2E$) APD. The multiplication region of this APD consists of an 80 nm-thick $Al_{0.2}Ga_{0.8}As$ layer sandwiched between two thin (~20 nm) layers of $Al_{0.6}Ga_{0.4}As$. The p and n layers on opposite sides of the multiplication region are heavily doped, resulting in a rather flat electric field profile across the multiplication region. The motivation for the $I^2E$ APD structure is to use heterostructures to help carriers impact ionize at preferred locations, resulting in lower excess noise. As shown in Ref. [14], photoelectrons diffuse through the p-region and are accelerated in the $Al_{0.6}Ga_{0.4}As$ layer in the multiplication region. Owing to the higher ionization threshold of $Al_{0.6}Ga_{0.4}As$, very few ionization events occur while the electrons gain energy. Once these electrons enter the $Al_{0.2}Ga_{0.8}As$ layer, they impact ionize in a more concordant manner. The spatial localization of the ionization events is the key to the more deterministic ionization behavior and the reduced excess noise in such an APD. Although these arguments explain the overall behavior of reduced noise, the measured <1 value of the excess noise factor is inconsistent with the definition of $F(<M>)$.

In order to understand the origin of the apparent conflict between theory and experiments, we need to re-examine the definition of the excess noise factor. In experiments, noise power, photocurrent and gain are measured. The excess noise factor can only be indirectly calculated using the assumption

$$F(<M>) = S_I(0)/(2e <M>^2 I_0), \qquad (1)$$

where $S_I(0)$ is the noise power at 0 Hz, and $I_0$ is the unity-gain photocurrent. In the following we review the origin of Eq. (1), following closely the derivations of van der Ziel [17].

The current power can be obtained using the Wiener-Khinchin theorem [17] by Fourier-transforming the current autocorrelation function, i.e., $S_I(f)=2F(<I(t)I(t+\tau)>)$. The DC component of $I(t)$, after autocorrelation and Fourier transform, becomes the signal power, $<I>^2\delta(f)$. We shall focus on the AC component, or the noise part of $S_I(f)$.

It can be proved that at $f\sim 0$ Hz [17],

$$S_n(0)=2 \ var\ n=2\ (<n^2> - <n>^2). \qquad (2)$$

If $n(t)$ is a Poisson process, $var\ n=<n>$. The electrical current $I(t)=en(t)$, where $e$ is the charge and $n(t)$ is the series of electron arrival events; therefore we have

$$S_I(0)=e^2S_n(0)=2\ e^2 var\ n = 2e^2<n>=2e<I>, \tag{3}$$

which is the well-known Schottky theorem for shot noise.

When multiplications are present, assuming $n$ electrons are initially photo-generated, each having a multiplication of $M_i$, the total number of collected electrons is $N=\sum_{i=1}^{n} M_i$. Under the assumptions $<M_i>=<M>$ *and* $<M_i^2>=<M^2>$, i.e, the multiplication and its variance are independent of the label "i" of the "i"th photoelectron, we have Burgess variance theorem [17]:

$$<N>=<n><M>,\ var\ N=<M>^2 var\ n + <n>\ var\ M, \tag{4}$$

consequently

$$S_I(0)=2\ e^2\ var\ N = 2\ e^2\ [<M>^2\ var\ n + <n>\ var\ M]. \tag{5}$$

If the photo-generated electrons are independent of each other, i.e., a Poisson distribution, we have $var\ n = <n>$. Using the definition of $var\ M$,

$$S_I(0)= 2\ e^2\ [< M>^2 <n> + <n>(<M^2>-<M>^2)]= 2\ e^2 <n><M^2>$$

$$=2e <M^2>/ <M>^2\ \ <M>^2 I_0\ = 2e\,F(<M>) <M>^2 I_0\,. \tag{6}$$

The validity of Eq. (6) depends critically on the assumptions for the Burgess theorem, which holds true only when the individual, multiplied current pulses resulting from each photoelectron have negligible width, i.e., "instantaneous amplification" [18]. The assumptions for Burgess theorem imply that each electron is associated with a certain photon (*i*). However, in the measurements of current and noise, photo-generated electrons and multiplication-generated electrons are indistinguishable when the current pulses overlap. When evaluating noise, the correlations among all the electrons should be considered, as we will do in the Monte Carlo simulations of the present work. The sum of all the electrons $N=\sum_{i=1}^{n} M_i$ should be over "time slices" instead over the label "i". Within a time slice the electrons could be the offspring of different photoelectrons. Burgess variance theorem in fact imposes a discrimination against the correlations within individual current pulses.

Recently developed "dead space" models explain lower noise in thinner APDs due to the change of the gain distribution [9, 10, 12, 13]. In these models, electrons need to travel a finite distance before gaining enough energy to impact ionize, hence the avalanche process is more deterministic than otherwise. The same results can be obtained in Monte Carlo simulations of gain distributions of individual electrons in a thin and a thick APD (Fig. 2). The dead-space models, and previous Monte Carlo models that evaluate $F(<M>)$ by counting the carriers, both neglect the importance of temporally finite-width pulses that are within themselves correlated, and such intra-pulse correlations contribute to the noise power as do the inter-pulse correlations. Hence, these models still assume Eqs. (4) to (6) to be true and predict the excess noise factor to be always ≥1. To clearly see this point, we consider a simple case where the ratio of hole-to-electron ionization coefficients $k=\beta/\alpha=0$. Traditional APD theory predicts $F(<M>)\rightarrow 2$ when $<M>\rightarrow \infty$, based on McIntyre's formulation [8],

$$F(<M>) = <M> [1-(1-k)(1-\frac{1}{<M>})^2]. \qquad (7)$$

One way to appreciate the role of dead space is to study an idealized APD where electrons can impact ionize continuously, i.e., no dead space, hence the gain $G=I_0\,exp(gL)$ along the length ($L$) of the device, where $g$ is the gain per unit length. In this case, the associated shot noise power can be written as

$$S(f) = 2eI_0 \exp(2gL) + \int_0^\infty 2e[I_0 \exp(gx)d(gx)]\exp[2g(L-x)]$$

$$= 2eI_0 \exp(2gL)[2 - \exp(-gL)]. \tag{8}$$

When $L \rightarrow \infty$, $S(f) = 2eI_0 G^2 \times 2$. The limiting case of $F(<M>) \rightarrow 2$ actually suggests that Eq. (7) only applies to thick APDs where the dead space length is negligible compared with the device thickness. Introducing the dead space will predict $F(<M>)$ to be less than the value predicted by Eq. (7) (but never <1).

If the definition of $F(<M>)$ is maintained, we should rewrite Eq. (1) and the experimentally measured excess noise factor has an effective value

$$\gamma F(<M>) = S_I(0)/(2e<M>^2 I_0), \tag{9}$$

where the coefficient $\gamma$ is closely related to the Fano factor. Although $\gamma=1$ when $<M>=1$, there is no reason to assume this is true when $<M>\neq 1$. We note that here $\gamma$ is treated as a numerical coefficient, which depends on $<M>$ in a complex way and will be evaluated using Monte Carlo simulations.

At low (~2) gains, the multiplication process may be sub-Poisson. An impact ionization event at an earlier time $t_1$ near the first heterojunction may prevent another impact ionization (from the same parent photoelectron) until after $t_2$. This is analogous to a traveling high-energy particle producing ion pairs along its way. Earlier ionizations lead to lower probability for ionizations later on. Such a negative correlation results in a Fano factor < 1 [19]. This happens when the energy budget is limited (or for an APD with a fixed value of gain, such as $<M>\sim 2$). We note that the negative correlation is less significant, if present at all, in homojunction APDs because the spatial locations of ionization events are more randomized [14]. At high gains, the multiplication process may become super-Poisson. That is, an ionization event at $t_1$ will lead to more and more impact ionization events later on, just as in the case of a secondary electron dynode [17]. This is only viable with an unlimited energy budget, which may be achieved for an APD by increasing the bias voltage.

In the Monte Carlo model described in Ref. [15], $F(<M>)$ was calculated using the statistics of output electrons and the results would be the same if the initial photo electrons were injected at the same time or injected continuously. Figures 3 and 4 show the electron ionization event distribution inside the APD and the current pulses, assuming the initial electrons are all injected at $t=0$. The electron ionization events form peaks near both heterojunctions, and the magnitudes of the peaks depend strongly on the average gain. Consequently, the pulse shapes vary with gain, implying different intra-pulse correlations at different gain values. At higher (~20) gains, the peak near the second heterojunction dominates [15] and the intra-pulse correlations may be less important.

To calculate noise spectra using the autocorrelation method, the photoelectrons must be injected in a span of time. We make the following assumptions in the present Monte Carlos simulation: (a) The initial photoelectrons are independent of each other in time and follow Poisson statistics. (b) We neglect the interactions among all the electrons, thus avoiding possible complications due to the correlations introduced by electron Coulomb interactions. The APD in this work has a thickness of ~100 nm and a diameter of ~160 μm, and is biased at ~10 V. In the low (<10) gain regime, the amplified current is in the order of micro-amps. This corresponds to a very low ($10^{11}$ $cm^{-3}$) electron density and the resulting electric field is negligible compared with the external field. (c) We neglect the quantum reflection or tunneling of carriers at heterojunctions. Due to the higher carrier energy (> 1eV) involved in APDs, the reflection and tunneling at the heterojunctions are not important [20]. Under these assumptions, any correlations in the simulated current result from impact ionizations.

The current is recorded as a function of time, using the Ramo-Shockley theorem $I(t) = \sum_i q_i v_i / L$, where $q_i$ is the charge of electrons or holes, $v_i$ is the carriers' instantaneous velocity, and $L$ is the thickness of the multiplication region. The autocorrelation function is calculated using [21]

$$C_I(j\Delta t) = \overline{I(t')I(t'+j\Delta t)} = \frac{1}{m'-m}\sum_{i=1}^{m'-m} I(i\Delta t)I[(i+j)\Delta t], \qquad (10)$$

where $m' = 1\text{x}10^6$, $m = 3\text{x}10^4$ and a time step $\Delta t$ = 50 fs are used.

We have simulated the scenario that $10^5$ electrons are photo-generated steadily in a large (compared with time scales of measurements and electron transport) time interval $T=50$ nanoseconds. Shot noise can be directly calculated from the autocorrelation function $<I(t)I(t+\tau)>$ of the resulting noisy, continuous current. The power spectra of the autocorrelation function, divided by $2eI_0 <M>^2$ and then normalized to the value for $<M>=1.0$ at 0 Hz, are plotted in Fig. 5. The overall shapes of these curves change drastically, and contain information about the carrier temporal correlations at various frequencies. In this Letter, we focus on the values near the 0-Hz end of the curves. These values in Fig. 5, denoted as $\gamma F(<M>)$, are plotted in Fig. 6. The 3σ error bars are calculated based on statistics of 16 independent sets of simulations with independent random number generators. It can be seen that $\gamma F(<M>)$ for gain values between 2 and 6 are firmly below 1. This qualitatively reproduces the measured excess noise factor in Fig. 1.

In conclusion, the spatial correlations of impact ionizations and the corresponding temporal correlations of amplified currents in an APD can result in a suppressed excess noise factor. This phenomenon is most profound in thin, heterostructure APDs where heterojunctions help localize ionization events, thus introducing strong correlations within current pulses. These correlations are gain-dependent because the localizations of impact ionizations depend on the gain values. Monte Carlo simulations reconcile the apparent conflict between the fact that some measured APD excess noise factors are less than 1 while mathematically it is required that $F(<M>) \geq 1$. The reason lies in the fact that what actually measured in experiments is $\gamma F(<M>)$ not $F(<M>)$. Finally, we note that this shot noise suppression mechanism could not be utilized to improve the signal-to-noise ratios of APDs, as the amplified current signal is within itself correlated for the same reasons.

The authors wish to thank Drs. John P. R. David, Majeed Hayat, Graham Rees, Bahaa Saleh, and Malvin Teich for helpful discussions. This work was supported by DARPA through the Center for Heterogeneously Integrated Photonics and the 3D Imaging Program.

Figure Captions:

Fig. 1 Comparison of measured excess noise factors and simulated *F(<M>)*. Open diamonds are measured data for a heterostructure APD described in Ref [14]. The simulations are based on a Monte Carlo model developed by Ma et al. [15]. The discrepancy between simulation and measurements for the heterostructure APD is apparent.

Fig. 2: Gain distribution for a 0.17-um-thick and a 1.44-um-thick GaAs homojunction APD from Monte Carlo simulations. The gain distribution of the thin APD does not have the high-gain tail as in the thick APD case, due to the “dead space” effect. This results in a smaller *F(<M>)* for thinner devices.

Fig. 3: Electron impact ionization events in the APD at various gains.

Fig. 4: Pulse responses at different gains.

Fig. 5: Normalized noise power spectra at various gains.

Fig. 6: Simulated excess noise factors for the heterostructure APD.

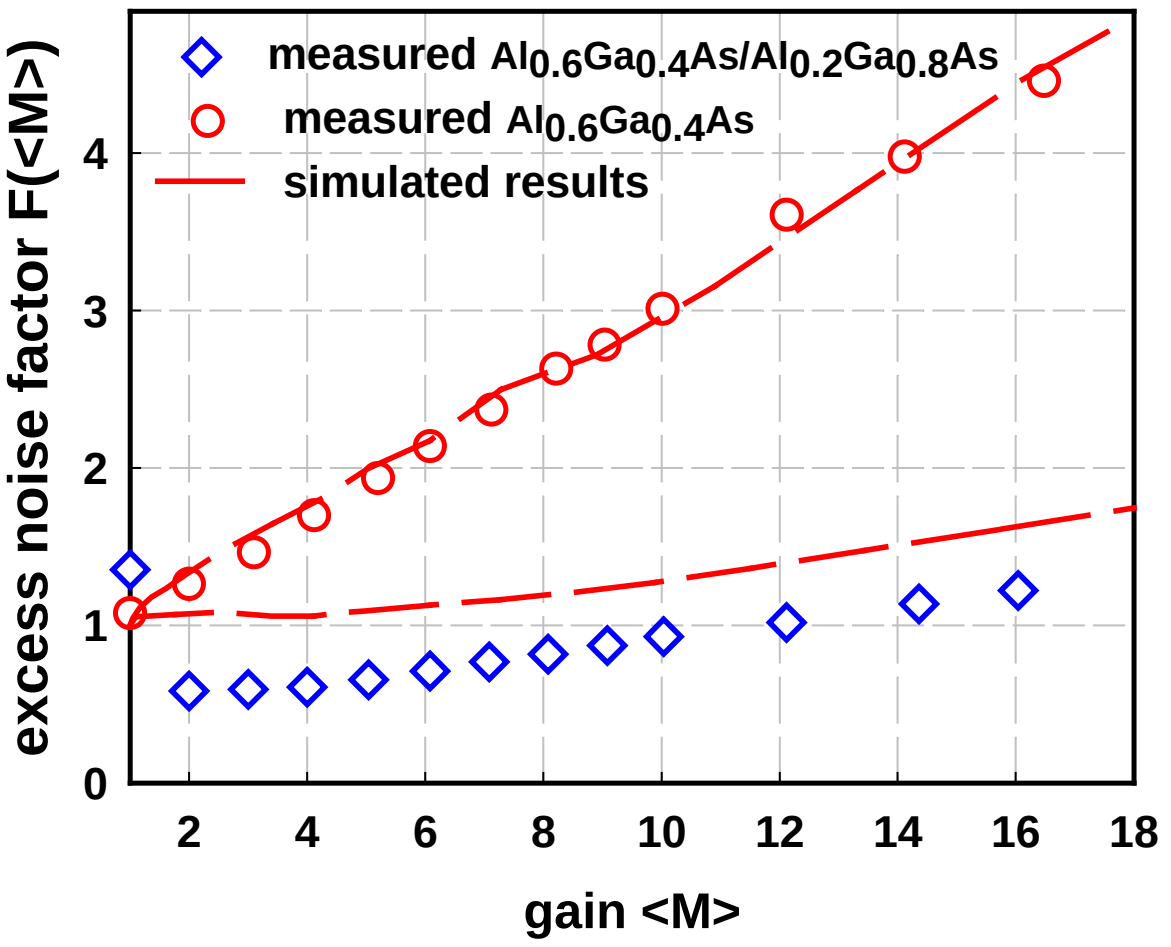

Fig. 1

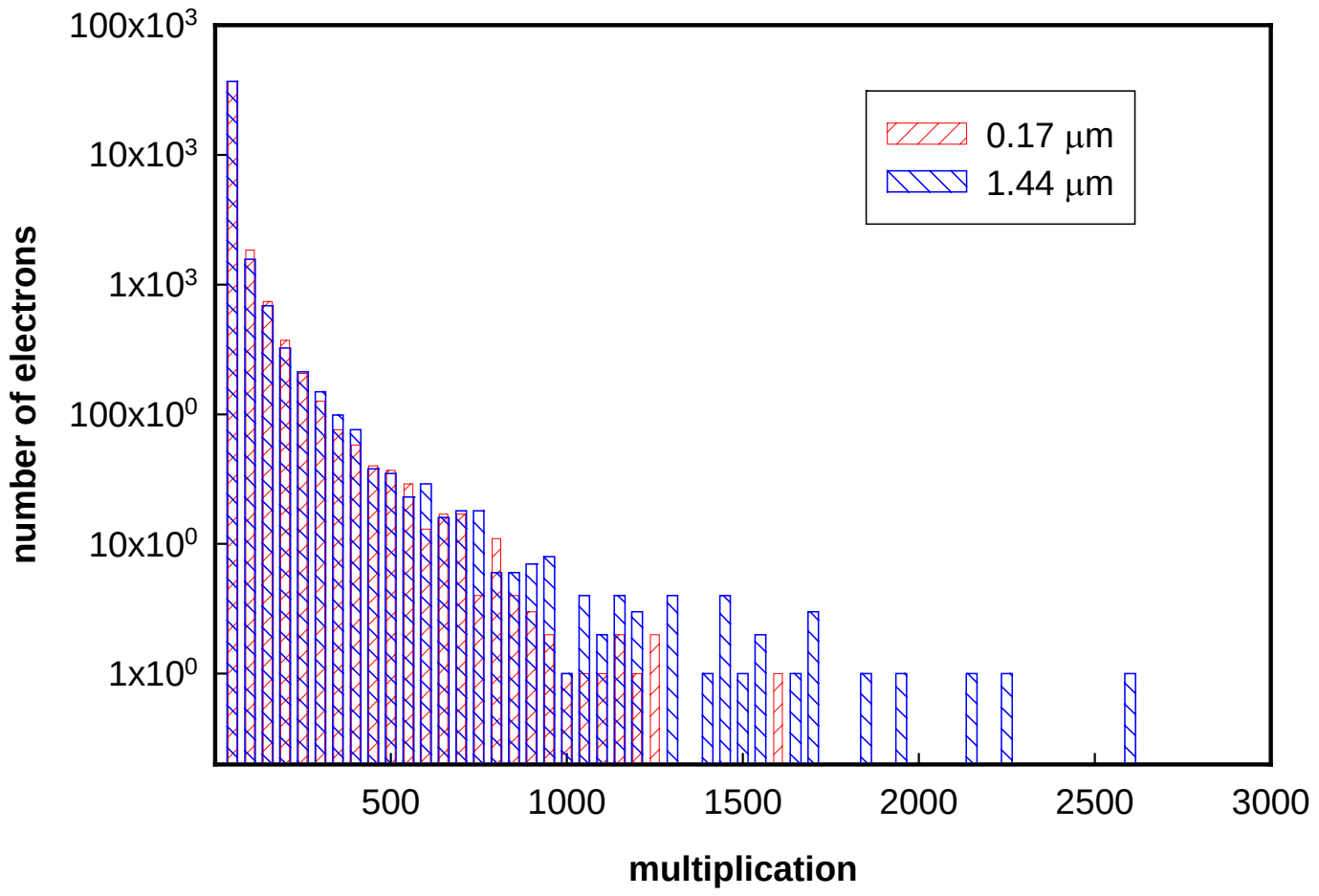

Fig. 2

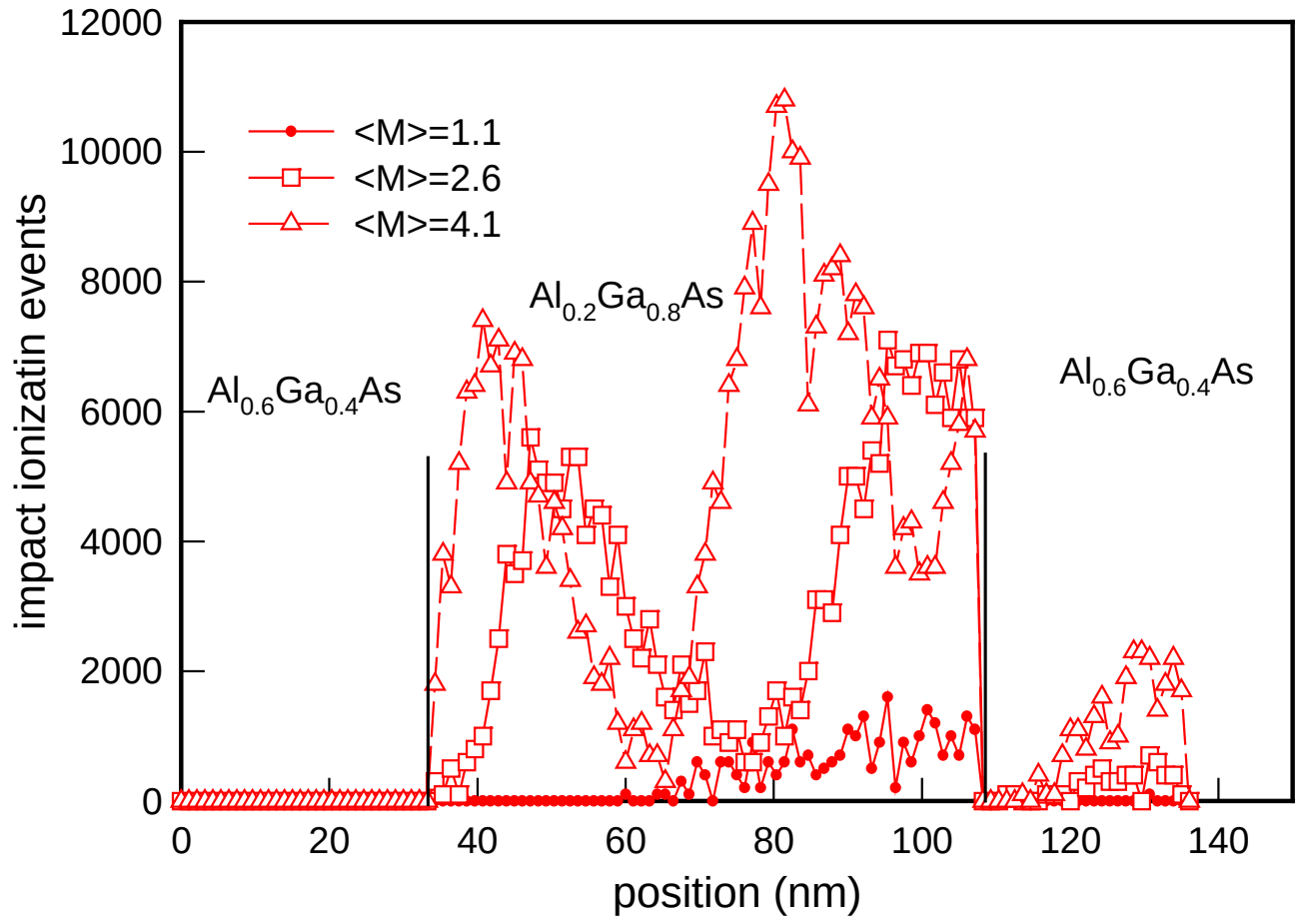

Fig. 3

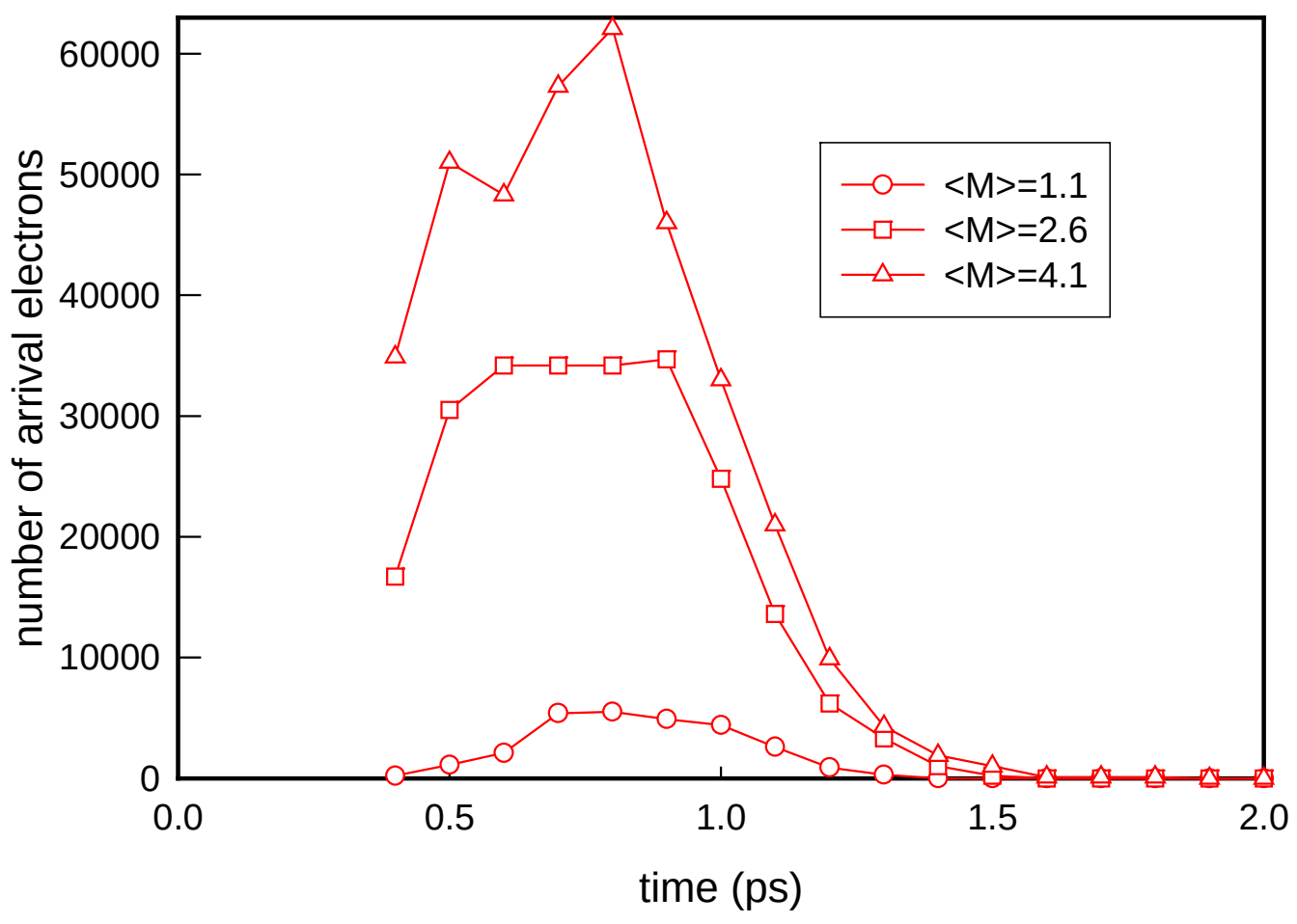

Fig. 4

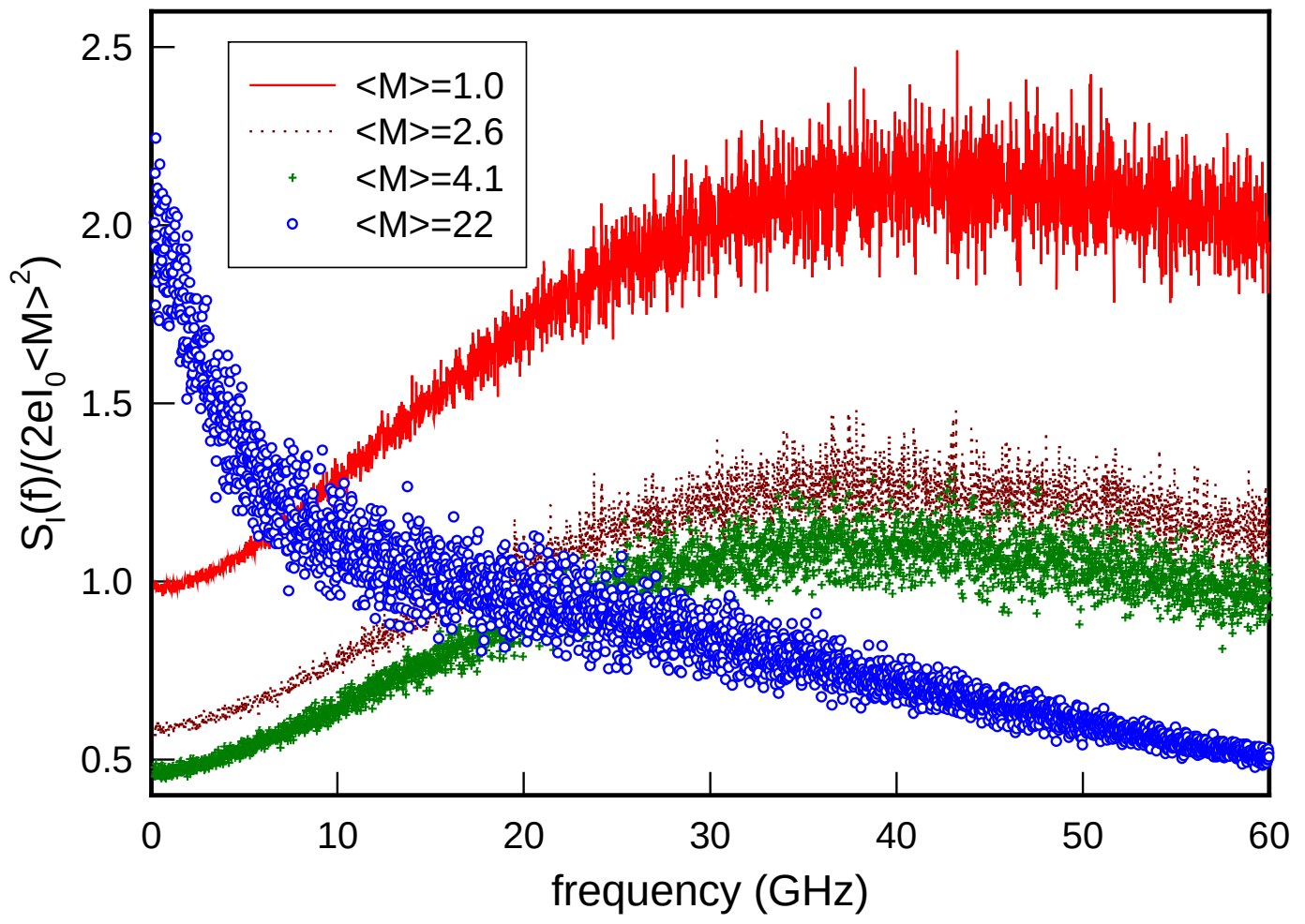

Fig. 5

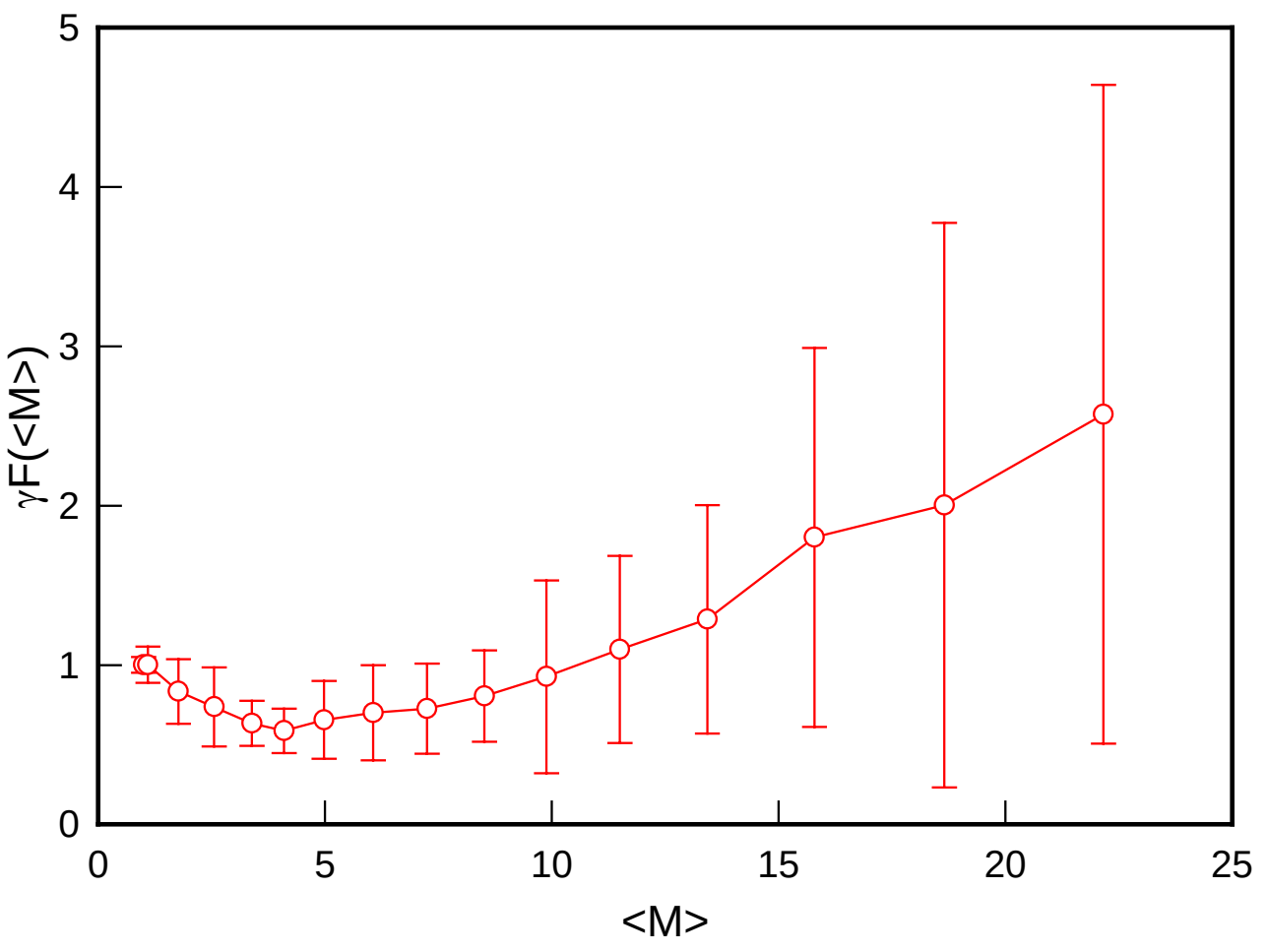

Fig. 6